\documentstyle[12pt]{article}
  \advance\oddsidemargin  by -1in
  \advance\evensidemargin by -1in
  \advance\topmargin      by -0.5in
\makeindex

\def\Pom{{\bf I\!P}}
\def\lsim{\mathrel{\rlap{\lower4pt\hbox{\hskip1pt$\sim$}}
    \raise1pt\hbox{$<$}}}         
\def\gsim{\mathrel{\rlap{\lower4pt\hbox{\hskip1pt$\sim$}}
    \raise1pt\hbox{$>$}}}         
\def\beq{\begin{equation}}
\def\eeq{\end{equation}}
\def\bea{\begin{eqnarray}}
\def\eea{\end{eqnarray}}

\begin{document}

\begin{flushright}
{\em ITEP-PH-5/98\\
FZ-IKP(TH)-1998-33}
\end{flushright}

\vspace{1.0cm}
\begin{center}
{\Large \bf  Precocious asymptopia for charm from  the running  BFKL\\

\vspace{1.0cm}}

{\large \bf N.N. Nikolaev$^{\alpha}$ and  V. R. Zoller$^{\beta}$}\\
\vspace{0.5cm}
$^{\alpha}${ \em
Institut  f\"ur Kernphysik, Forschungszentrum J\"ulich,\\
D-52425 J\"ulich, Germany\\
E-mail: kph154@ikp301.ikp.kfa-juelich.de}\\

$^{\beta}${\em Institute for  Theoretical and Experimental Physics,\\
Moscow 117218, Russia\\
E-mail: zoller@heron.itep.ru}\\

\vspace{0.5cm}
{\bf Abstract}
\end{center}
The running BFKL equation  gives rise to a series of
 moving poles in the complex $j$-plane. Corresponding eigenfunctions 
(color dipole cross sections) are the oscillating functions of the color 
dipole size $r$. 
 The  first nodes for all sub-leading solutions
(color dipole cross sections) accumulate at
$r_1\sim 0.1\,{\rm fm}$.
Therefore the processes dominated by the dipole sizes $r\sim r_1 $
 are free of sub-leading BFKL corrections. A practically important example -
 the leptoproduction of charm. In a wide range of $Q^2$ the calculated
$F_2^{cc}(x,Q^2)$ is exhausted by the leading BFKL pole and
gives a perfect description of the experimental data.\\

\vspace{0.5cm}

The generalized Balitskii-Fadin-Kuraev-Lipatov (BFKL \cite{BFKL})
 equation for the interaction
cross section $\sigma(x,r)$ of the color dipole $\vec r$ with the target
reads \cite{PISMA1}

\beq
 {\partial \sigma(x,r)\over \partial \log(1/x) } ={\cal K}\otimes
\sigma(x,r)
\label{eq:BFKL}
\eeq
where  $x$ is the Bjorken variable.
 The kernel ${\cal K}$ is related to the flux of the Weizs\"acker-Williams
 soft gluons $|\vec E(\vec\rho_1)-\vec E(\vec\rho_2)|^2$.
 The Asymptotic Freedom (AF)  dictates
that the  chromoelectric fields $\vec E(\vec\rho)$ be calculated with the running QCD
charge $g_S(r_m)=\sqrt{4\pi\alpha_S(r_m)}$ taken at shortest relevant distance
$r_m=$min$\{r,\rho\}$ and
$$\vec E(\vec\rho)=g_S(r_m){{\vec \rho}/\rho^2}\times
({\rm screening\, factor}).$$
 Within the infrared
 regularization scheme described in \cite{PISMA1,NZZJETP,PLNZ1,PLNZ2}

\beq
 \vec E(\vec\rho)=g_S(r_m){{\vec \rho}\over\rho R_c}K_1(\rho/R_c)\,,
\label{eq:EVEC}
\eeq
  where $ K_1(x)$ is the modified Bessel function.
 Our numerical results
 are for the Yukawa screening radius  $R_c= 0.27\,{\rm fm}$.
The anlysis of  the lattice QCD data on
 the field strength correlators suggests similar $R_c$ \cite{MEGGI}. 
 The so  introduced  running coupling may
not exhaust all NLO effects but it correctly describes the crucial
enhancement of long distance, and suppression of short distance, effects
 by AF.

Our findings on   the running  BFKL equation which are
 of  prime importance for the problem under discussion 
 are as follows \cite{DIS97,JETP}.
The spectrum of the running BFKL equation is  a series of
 moving poles $\Pom_n$ in the complex $j$-plane with eigenfunctions
\beq
\sigma_{n}(x,r)=\sigma_{n}(r)\exp(\Delta_{n}\log(1/x) )
\label{eq:SIGMAN}
\eeq
  being a solution of
\beq
{\cal K}\otimes \sigma_{n}=\Delta_{n}\sigma_{n}(r)\,.
\label{eq:EIGEN}
\eeq

  The leading eigen-function $\sigma_0(r)$
 is node
free.  The  sub-leading  $\sigma_n(r)$
 has $n$ nodes.
 The intercepts
$\Delta_n$
 closely, to better than $10\%$, follow the law

\beq
\Delta_n= {\Delta_0\over (n+1)}
\label{eq:EIGVAL}
\eeq
suggested earlier by Lipatov \cite{LIPAT86} .
The intercept of the leading pole trajectory,
with our specific choice of the infrared regulator,
 $R_c=0.27\, {\rm fm}$, is $\Delta_0\equiv\Delta_{\Pom}=0.4$.
The sub-leading $\sigma_{n}(r)$ represented  
in term of  ${\cal E}(r)= \sigma_n(r)/r$, to a crude 
approximation is similar to Lipatov's quasi-classical eigenfunctions
 \cite{LIPAT86},
$${\cal E}_n(r)\sim \cos[\phi(r)]\,.$$
With  $R_c=0.27\, {\rm fm}$
  the  node of $\sigma_{1}(r)$
is located at $r=r_1\simeq 0.05-0.06\,{\rm fm}$,
for larger $n$ the first node
moves to a somewhat  larger $r\sim 0.1\, {\rm fm}$.
 Hence,
 $\sigma(x,r_1)$ is dominated by $\sigma_0(x,r_1)$. 
This observation explains
 the precocious asymptopia
for the dipole cross section, 
$$\sigma(x,r) \propto (1/x)^{\Delta_{\Pom}}\,,$$ 
at $r\sim 0.1\, {\rm fm}$ derived previously from the numerical studies
of the running BFKL equation \cite{NZZJETP, PLNZ1,PLNZ2}.
Consequently, zooming at   $\sigma(x,r_1)$ one can readily
 measure  $\Delta_{\Pom}$. The point we want to make here is that
because $r_1\sim 1/m_c$, the excitation of open charm provides the desired
zooming. Indeed, we shall demonstrate  that  
  the effect of suppression of the sub-leading BFKL terms in $F_2^{cc}(x,Q^2)$ 
 is  remarkably strong{\footnote {The preliminary results have
 been reported at the DIS'98 Workshop
\cite{DIS98}}}.\\

The color dipole representation for the  charm structure
 function (SF)   reads \cite{NZ91} 
\beq
F_2^{cc}(x,Q^2)={Q^2\over {4\pi\alpha_{em}}}\langle\sigma\rangle=
{Q^2\over {4\pi\alpha_{em}}}\int_0^1 dz\int
d^2\vec r|\Psi^{cc}(z,r)|^2\sigma(x,r)\,,
\label{eq:F2CC}
\eeq

Starting with the BFKL-Regge expansion
\beq
\sigma(x,r)=\sigma_0(r)(x_0/x)^{\Delta_0}+\sigma_1(r)(x_0/x)^{\Delta_1}+
\sigma_2(r)(x_0/x)^{\Delta_2}+...\,.
\label{eq:sigma}
\eeq
with $\Delta_n$ determined in \cite{DIS97,JETP}
we arrive at the BFKL-Regge expansion for the charm SF
\beq
F_2^{cc}(x,Q^2)=\sum_nf_n^{cc}(Q^2)(x_0/x)^{\Delta_n}\,,
\label{eq:F2BFKL}
\eeq
where the charm eigen-SF is as follows
\beq
f_n^{cc}(Q^2)={Q^2\over {4\pi\alpha_{em}}}\langle\sigma_n\rangle \,.
\label{eq:fNBFKL}
\eeq

In conjunction with the explicit form
of the $c\bar c$ light-cone wave function,
$\Psi^{cc}(z,r)$ \cite{NZ91},
\beq
|\Psi_T^{cc}(z,r)|^2={24\alpha_{em}\over 9(2\pi)^2}
\left\{\left[z^2+(1-z)^2\right]\varepsilon^2K_1(\varepsilon r)^2
+m_c^2K_0(\varepsilon r)^2\right\},
\label{eq:PSICC}    
\eeq
where $K_{0,1}(x)$ are the modified Bessel functions,
 $\varepsilon^2=z(1-z)Q^2+m_c^2$,
$m_c=1.5\,{\rm GeV}$ is the $c$-quark mass and $z$
 is the light-cone fraction of photon's 
momentum carried by the quark of the $c\bar c$ pair,
 the eqs.(\ref{eq:F2CC}), (\ref{eq:fNBFKL})
 show that the integral over $r$ in (\ref{eq:F2CC})
is dominated by 

$$Q^{-2}\lsim r^2\lsim m_c^{-2}.$$

Indeed, making use of the properties of  modified Bessel functions,
after z-integration one can write
\beq
f_n^{cc}(Q^2)\propto \int_{1/Q^2}^{1/m_c^2}{dr^2\over r^2}{\sigma_n(r)\over r^2}\,.
\label{eq:FNINT}
\eeq
The dipole cross section $\sigma_n(r)$ in (\ref{eq:FNINT})
 is an oscillating function of $r$ with
the first node located inside   the integration region.
 Then
 in  a broad range of $Q^2$
one has strong cancellations   in   (\ref{eq:FNINT})
 for  sub-leading poles 
 which result in 
the leading pole dominance in charm production (Fig. 1). 
For large $Q^2$, far  beyond the nodal region, 
the effect of cancellations disappears
and
\beq
f_n^{cc}(Q^2)
\propto 
\left[{ \alpha_S(Q^2)}\right]^{-\gamma_n} 
\label{eq:F2ASS}
\eeq
with $\gamma_n={4\over 3\Delta_n}$.

For practical purposes it is convenient to represent $f_n^{cc}(Q^2)$
 in an analytical  form. The  parameterization for the leading pole SF
reads
\beq
f_0(Q^2)=
a_0{R_0^2Q^2\over{1+ R_0^2Q^2 }}
\left[1+c_0\log(1+r_0^2Q^2)\right]^{\gamma_0}\,,
\label{eq:F0}
\eeq
where
$\gamma_0={4\over 3\Delta_0}\,.$

For $n\geq 1$ the functions $f_n(Q^2)$ can be approximated by
\beq
f_n(Q^2)=a_nf_0(Q^2){1+R_0^2Q^2\over{1+ R_n^2Q^2 }}
\prod ^{n}_{i=1}\left(1-{z\over z^{(i)}_n}\right)\,,
\label{eq:FNQ2}
\eeq
where
\beq
z=\left[1+c_n\log(1+r_n^2Q^2)\right]^{\gamma_n}-1
\label{eq:Z}
\eeq
and
\beq
\gamma_n=\gamma_0 \delta_n 
\label{eq:GAMMA}
\eeq
with parameters  listed in the Table.

\vspace{5mm}

\begin{tabular}{|l|l|l|l|l|l|l|l|} \hline
$n$& $a_n$ &  $c_n$ & $r_n^2\,,$ ${\rm GeV^{-2}}$ & 
$ R_n^2\,,$ ${\rm GeV^{-2}}$ &$z^{(1)}_n$ & $z^{(2)}_n$ & $\delta_n$ \\ \cline{1-8}
0  &  0.0214      &  0.2619 & 0.3239 & 0.2846 &    &  &  \\ \cline{1-8}
1  & 0.0782& 0.0352&0.0793&0.2958&0.2499&  & 1.9249\\ \cline{1-8}
2&0.0044 &0.0362&0.0884&0.2896&0.0175 &3.447&1.7985\\ \cline{1-8}
\end{tabular} 
\vspace{5mm}

 In Fig.2
our predictions for the charm structure function are compared
with data from H1 \cite{H1cc} and ZEUS \cite{ZEUScc}.  We correct for
threshold effects  by the rescaling \cite{Barone} 
$x\to x(1+4m_c^2/Q^2)$.
From both Fig.1 and Fig.2 it follows
that
the charm production in a wide range of the photon virtualities,
 $Q^2\lsim 10^2\,{\rm GeV^2}$,
 provides the unique opportunity of getting hold of elusive BFKL
asymptotics and measuring $\Delta_{\Pom}$
 already  at currently available $x$.


\newpage

{\bf Figure captions}
\begin{enumerate}
\item[{\bf Fig.1}]
Modulus of the  charm  eigen-SF $|f^{cc}_n(Q^2)|$
for the BFKL poles  with $n=0,1,2\, $.
\item[{\bf Fig.2}]
The predicted  charm structure function $F^{cc}_2(x,Q^2)$ (solid line) vs.
 H1   and ZEUS   data. The contribution of the leading pole with
$\Delta_{\Pom}=0.4$ is shown by dashed line. Fig.2a corresponds to
 $Q^2=45\,GeV^2$, Fig.2b - $Q^2=25\,GeV^2$, Fig.2c - $Q^2=12\,GeV^2$ and 
Fig.2d - $Q^2=7\,GeV^2$.
\end{enumerate}

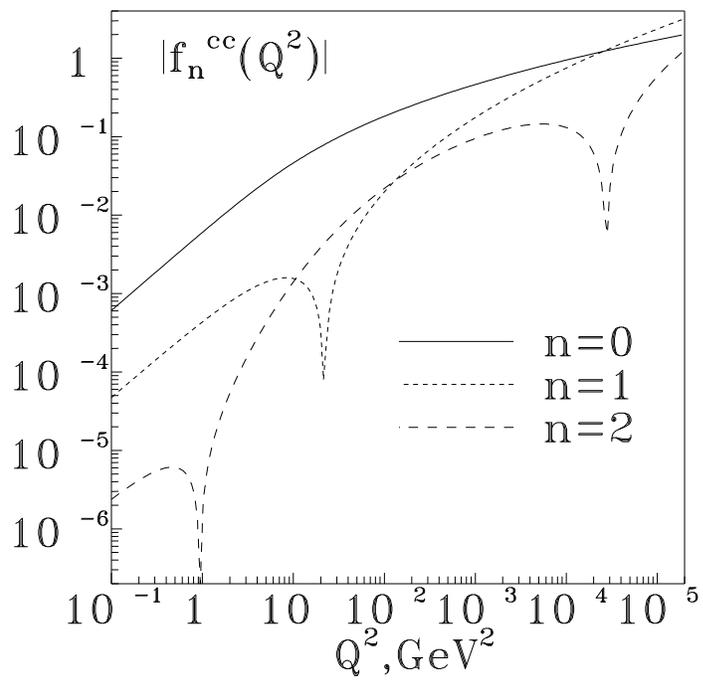
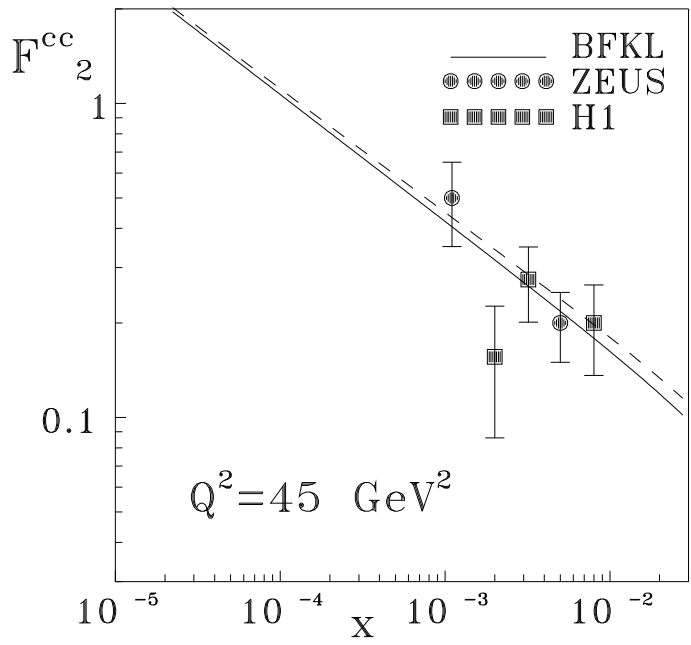
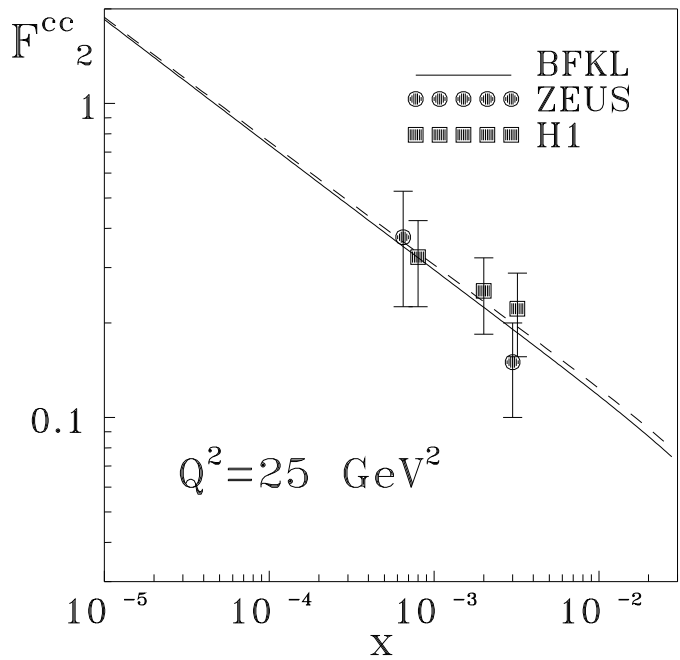
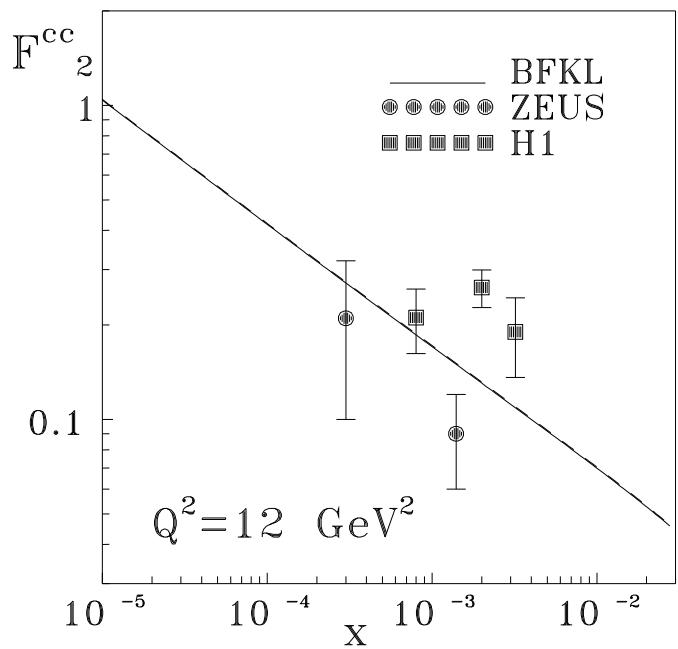
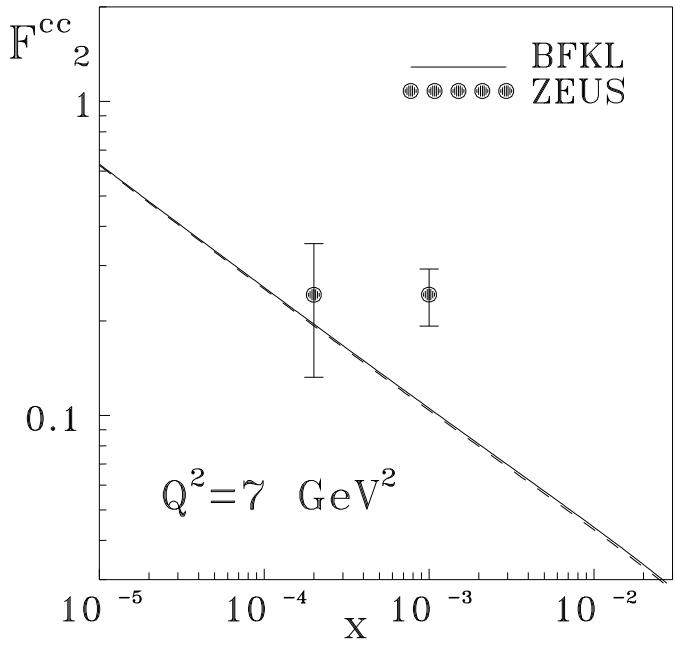
\end{document}